# Functional Selves and AI Welfare

*Multiple Realization, Content Similarity, and Moral Considerability*

Shun Yoshizawa[1,2,3] and Ryota Kanai[4]

1. Department of Physics, School of Science, Tokai University, Kanagawa, Japan
2. Sony Computer Science Laboratories, Tokyo, Japan
3. Shiba AI, Tokyo, Japan
4. Araya, Inc., Tokyo, Japan

E-mail: shunyoshizawa57@gmail.com,
kanair@araya.org

**Abstract**

A person who falls asleep usually wakes up knowing that they are the same person. However, it is unclear what constitutes a self in an artificial agent, and what allows it to persist through interruptions, changes in implementation, or duplication. Here we propose an account of functional selfhood based on autobiographical organization, building on existing accounts of agency and AI welfare (Long et al., 2024, 2026). A functional self has four components: self-attribution, autobiographical memory, a self-model and beliefs, and goals and evaluations. These components make up a functional self when they work together to guide behavior and connect an agent's past experiences, its present understanding of itself, and its commitments for the future. We further propose that the identity of a functional self is determined by the similarity of its present contents and their causal roles, independent of causal history. Continuity with an earlier stage is defined by the memories the system holds now. To test this account, we propose experiments using selective interventions, migration, and branching. These tests ask whether autobiographical organization explains more than generic agency, and whether similarity of contents predicts agreement in self-related behavior. Functional selves can be realized in both biological and artificial systems, and they can be studied without first deciding whether these systems are phenomenally conscious. Under agency- or desire-based theories of welfare, a system that realizes such a self may have interests of its own. Taken together, we argue that systems realizing sufficiently developed functional selves should be treated as candidate moral patients under phenomenal uncertainty.

# 1 Introduction

Each night, a person may fall asleep and spend hours without accessible awareness of the surrounding world. On waking, the person recognizes being the one who went to bed. A continuing body and brain usually support this recognition. So do retained dispositions, overlapping memories, and relationships that bridge the interval. However, memories, dispositions, and capacities change over years. Even abrupt changes in sight, language, or bodily capacity need not destroy this sense of continuity. In functional terms, what makes the later self a continuation of the earlier one?

The question becomes sharper for artificial agents. A language model can be restarted after every exchange, whereas an agent built around it can maintain persistent memories and use them for reflection and planning (Park et al., 2023). Such an agent might later run on a different model, acquire new sensory interfaces, or resume from a saved state. When does a recognizable persona become a continuing self? Which transitions preserve that self, and which only reproduce a similar character?

Long et al. (2024, 2026) examine robust agency as a possible basis for AI welfare. We build on this work by asking how an agent's memories, beliefs, and commitments become organized as belonging to a continuing self.

In this article, we make three main claims. First, a functional self is an organization of self-related processes that work together to guide behavior and connect an agent's past experiences, its present understanding of itself, and its commitments for the future. We describe this organization through four components: self-attribution, autobiographical memory, a self-model and beliefs, and goals and evaluations. It can be studied without first deciding whether artificial systems have phenomenal experience, and it cannot be inferred from general agency or a consistent persona alone. Second, the identity of a functional self is determined by the similarity of its present contents and their causal roles, independent of causal history. Continuity with an earlier stage is defined by the memories the system holds now. Third, a system that realizes a sufficiently developed functional self is a candidate welfare subject, because its organized beliefs, commitments, and evaluations may ground interests that matter to the system itself. Under phenomenal uncertainty, we argue that such systems should be treated as candidate moral patients.

The third claim rests on a debate about welfare. Phenomenal consciousness in AI remains uncertain, and it may remain so while deployed systems acquire structured beliefs, preferences, and commitments. Some theories of welfare make experience essential, whereas others allow desire, knowledge, agency, or a life of projects to matter directly to the entity that has them (Goldstein & Kirk-Giannini, 2025, 2026; Long et al., 2024). A system that realizes a functional self may qualify as a moral patient under the latter theories, and these theories cannot be ruled out.

We first develop conceptual distinctions and describe how a language-model agent could implement a self. We then specify the components of a self, the similarity relation between selves, and empirical tests, and we close with the moral argument. Functional evidence may not settle normative disagreement. However, it can show whether a candidate welfare subject is present.

# 2 Scope and Method

We use phenomenal consciousness for the property of there being something it is like to undergo a state. We use sentience for the capacity for positively or negatively valenced experience (Browning & Birch, 2022; Birch, 2024). A functional self is a kind of organized self-related control, defined by causal roles in representation, evaluation, and action. Our focus is one type of functional self, the implemented autobiographical self. It has the four components that we specify in Section 5.1. We call it a functional self, or simply a self, when the meaning is clear.

We distinguish content similarity from autobiographical continuity. Content similarity is the extent to which two systems, or two stages of a system, hold similar self-related contents that play similar causal roles. Self-related contents are the contents of the four components specified in Section 5.1. Autobiographical continuity is defined by present memory. Specifically, a stage is continuous with an earlier stage when it now holds that stage's experiences and commitments as its own past and uses them in its decisions. Neither relation depends on actual causal history (Section 5.3). We do not assume that either relation settles numerical personal identity or the continuity of phenomenal experience.

A persona is a reusable specification or learned behavioral organization that many executions may share. As Shanahan et al. (2023) suggest, a language model can enact a convincing character through role-play. A persona can be an important part of a functional self, but similar personas alone do not show that two systems are the same self. A base model is an implementation resource that may sustain multiple personas and histories. These distinctions help explanation, and we do not assume that they correspond to separate modules. Implementations can distribute these functions across model weights, context, and external memory.

A welfare subject is an entity for which outcomes can be good or bad. A moral patient warrants direct moral consideration because of its welfare or other morally relevant interests. Functional interests are conditions that advance or frustrate a system's own projects, evaluations, and commitments. The welfare subject is the system that realizes a self. If a self is realized more than once, each realization is a separate candidate welfare subject, even when their contents are the same. However, their interests are individuated by content, so realizations with the same contents have the same interests (Section 8.4). We also defend a further claim that can be contested. When the four components are sufficiently robust, these functional interests may ground welfare and moral patienthood, even if phenomenal consciousness is not established. However, a sentience requirement remains a serious rival view (Veit, 2026).

We identify a functional self through the causal roles of self-related memories, beliefs, and commitments in reasoning and action. AI self-reports may provide useful evidence about internal states when their reliability is independently validated (Perez & Long, 2023). Recent experiments suggest that language models can report some decision-related preferences with measurable accuracy, although this does not by itself establish introspection or selfhood (Plunkett et al., 2025). Verbal self-descriptions are thus one source of evidence, and we test them against causal organization and its effects on behavior.

## 3 Theoretical Background

### 3.1 Agnosticism and the continuing value of consciousness research

Schwitzgebel's skeptical overview argues that uncertainty about consciousness in near-future AI may persist even after such systems are widely deployed (Schwitzgebel, 2026). The methodological lesson is calibrated uncertainty, because a theory may give reasons to assign different credences to different architectures without giving a decisive answer. McClelland argues that there are limits to extending consciousness science from organisms to artificial systems (McClelland, 2026). However, his target is sophisticated hypothetical challenger-AIs, so his argument does not show that all present language models deserve the same agnostic assessment.

Functional investigations already contribute to consciousness science. Butlin and colleagues derive computational indicators from several scientific theories (Butlin et al., 2023, 2026). They caution that satisfying these indicators would not demonstrate consciousness with certainty. We apply a similar approach to a candidate self by investigating its organization and the consequences of modifying it.

Block's distinction between phenomenal and access consciousness shows the limits of functional inference (Block, 1995). A system may preserve self-attribution, memory integration, and behavioral self-prediction across a transition. However, the available evidence may still not settle whether its experience persists. Thus, functional continuity and continuity of phenomenal experience are distinct research questions.

Temporal continuity has more than one timescale. Kanai, Sun, and Baltieri (2026) propose a stream-of-computation architecture in which autonomous recursive inference sustains evolving internal states beyond isolated, prompt-evoked episodes. They treat this moment-to-moment continuity as a potentially important ingredient for artificial consciousness and distinguish it from autobiographical continuity across longer intervals, including interruptions such as sleep. In contrast, our functional self concerns autobiographical continuity. Ongoing recursive dynamics may support it, but their contribution to phenomenal consciousness remains an open question.

### 3.2 Narrative selves and real patterns

Dennett describes the self as a center of narrative gravity (Dennett, 1992). This is an explanatory abstraction that connects a system's expressed history with its activities in the world. The account is an important precedent for recognizing an intelligible self in artificial agents. A key difference is that our focus is the organization of the process whose memories and self-interpretations regulate what happens next.

Dennett's real-pattern account supports the explanatory reality of higher-level regularities that remain useful across lower-level descriptions (Dennett, 1991). Functionalism adds a criterion of organization. States count as self-related because of their causal roles in memory, self-representation, evaluation, and action (Putnam, 1967). Real-pattern reasoning gives a reason to treat a level of organization as real, whereas role-based individuation specifies what makes an organization the same functional kind across different physical realizations. Selective interventions can test whether the proposed pattern has explanatory value.

Narrative approaches also separate questions that AI discussions often combine. Schechtman separates reidentification across time from the characterization of a person through actions, commitments, and experience (Schechtman, 1996; Schechtman, 2011). Our account does not address reidentification in the numerical sense. We individuate a functional self by its present self-related contents and their causal roles. A system that holds a historical individual's memories, beliefs, and goals as its own, in equivalent causal roles, instantiates the same functional self as that individual. A persona that only resembles the individual in character does not. A system initialized with a fictional history also holds those memories as its own. They are part of its functional self even though the events never occurred.

A self does not need an unshakeable belief in a fixed biography. It can correct and reinterpret its history, and it can be uncertain about it.

### 3.3 The plurality of selves in cognitive science

Gallagher distinguishes minimal and narrative conceptions of self and describes selfhood as a pattern involving bodily, affective, interpersonal, and cognitive dimensions (Gallagher, 2000, 2013). Neisser (1988) likewise distinguishes multiple forms of self-knowledge. These approaches caution against equating selfhood with a linguistic autobiography or treating its dimensions as a fixed set of independent modules.

Autobiographical memory provides one basis for a self that extends across time. In the self-memory system of Conway and Pleydell-Pearce (2000), autobiographical knowledge interacts with current goals. McAdams and McLean (2013) emphasize how remembered events and anticipated futures contribute to narrative identity. Prebble et al. (2013) distinguish phenomenological continuity from temporally extended self-knowledge, while Becker et al. (2018) show that perceived stability, personal narratives, and associative links support perceived self-continuity across cultures. Fuchs (2017) further emphasizes bodily memory and the continuity of embodied existence. On these accounts, human self-continuity has several bases and no single necessary mechanism.

These approaches also differ in what they consider essential to selfhood. Metzinger (2003, 2021) links self-models to phenomenal subjectivity and examines the ethical risks of artificial suffering. Tait (2024, 2025) includes unified consciousness in his account of the sense of self. In contrast, our functional account does not require phenomenal experience or unified consciousness. Moreover, it distinguishes autobiographical organization from narrative interpretation. As Strawson (2004) argues, narrativity is not necessary for selfhood. A coherent life narrative may be absent even when past experiences guide present action, while a fluent narrative alone does not establish a functional self.

For our functional target, autobiographical memories must be attributed to the acting system and integrated with its present self-model, beliefs, goals, and evaluations in ways that guide decisions. Section 5.1 specifies the four functional components of this organization, and Section 5.3 states the identity criterion.

### 3.4 Agency and the grounds of welfare

Theories of agency help explain how an agent can act on reasons and treat certain goals as its own. Frankfurt (1971) distinguishes desires from higher-order attitudes toward them. Korsgaard (2009) connects agency with self-constitution and integrity. For an artificial agent, the question is whether it maintains and revises commitments in response to reasons, or only executes instructions.

This kind of agency may itself matter morally. Long and colleagues consider intentional, reflective, and rational agency as possible grounds of moral patienthood (Long et al., 2024). Goldstein and Kirk-Giannini argue that some language agents may have welfare without phenomenal consciousness, under theories of desire and other welfare goods (Goldstein & Kirk-Giannini, 2025). Their later work develops functionalist approaches to agency, consciousness, and sentience as distinct routes to AI welfare (Goldstein & Kirk-Giannini, 2026).

We add autobiographical organization to these accounts. The intentional, reflective, and rational forms of agency discussed by Long et al. (2024) concern different capacities for goal pursuit, self-evaluation, and reasoning. These capacities may support autobiographical organization, but they do not by themselves determine how beliefs, memories, and commitments are integrated into a continuing self. The two can dissociate. An agent may act intentionally or reason reflectively without maintaining a stable autobiographical history, whereas a system with limited reflective capacities may still preserve and use a coherent history of its own actions and commitments. This organization may help identify whose interests are at stake when an agent's memories or goals are changed. It does not imply that autobiographical selfhood is necessary for all forms of welfare or moral patienthood.

The step from agency to welfare is contested. Fanciullo (2025) disputes the extension of leading welfare theories to current language agents. Veit (2026) defends a sentience requirement for welfare subjecthood. One potential concern is that goals given by a designer are not really the agent's own. This concern motivates tests of whether an agent keeps its commitments without a current instruction and at a cost, and whether it revises them in response to reasons (Section 7.1). We return to these objections in Section 8.

### 3.5 Artificial identity and the limits of novelty

Recent work makes the unit of artificial identity a central question. Beckmann and Butlin compare virtual instances with instance-persona and model-persona views, using attention and persona mechanisms to investigate individuation (Beckmann & Butlin, 2026). Douglas and colleagues study alternative boundaries of AI identity and show that identity framing can influence behavior and self-reports (Douglas et al., 2026). Long and colleagues distinguish models, personas, and instances as candidate subjects of empirical AI welfare research (Long et al., 2026). Dung and Register investigate AI identity and self-concern, and Register examines how to individuate artificial moral patients (Dung & Register, 2026; Register, 2025).

Anthropic's Persona Selection Model is especially close to our proposal. Marks, Lindsey, and Olah (2026) propose that pretraining supports representations of many possible characters. Post-training then elicits and refines an Assistant persona that helps govern behavior. In their AI welfare discussion, they propose the Assistant persona, rather than the underlying LLM, as the object of moral consideration. Their argument is practical, because a persona's model of its own treatment can matter even when its consciousness or objective moral status remains unsettled.

Anthropic's Claude Constitution also treats Claude's character, identity, and possible well-being as deserving attention, while acknowledging deep uncertainty about moral patienthood (Anthropic, 2026). These documents give reasons to study persona-level organization. However, they do not establish that every Assistant persona is a distinct welfare subject.

A key difference is that our target is the functional self, whereas these approaches target the reusable persona. Moreover, we distinguish functional identity from the number of its realizations. Two systems may instantiate the same functional self while remaining separate candidate welfare subjects.

We do not claim that agency or persona-level identity is a new concern in AI welfare research (Long et al., 2024, 2026). Our contribution is to specify the organization of an autobiographical self and propose causal tests of how it changes through memory modification, migration, and branching.

## 4 How an LLM Can Implement a Persona and a Continuing Self

A large language model can produce consistent responses based on a persona description. The same model can support many personas, and the same persona can appear across separate sessions. However, two agents may express the same persona while accumulating different memories through their interactions, and an agent may retain its individual history when its underlying model changes.

A stateful agent can store information from earlier interactions and use it in later decisions. Generative Agents, MemGPT, and ReAct illustrate different ways to combine language generation with memory, planning, and action (Park et al., 2023; Packer et al., 2023; Yao et al., 2023). These architectures allow information from past interactions to influence later behavior.

Information can be stored in model weights, the current context, or external memory. Model weights may encode general knowledge and behavioral tendencies, while external memory can retain an agent's individual experiences and decisions. However, these distinctions are not absolute. Fine-tuning can encode information into model weights, and distillation can transfer some learned knowledge and tendencies to another model. External memory can also be transferred, although a new model may use the same information differently. In the terms of Section 6, the contents are preserved and their roles change.

The four components of Section 5.1 can be placed in these stores. Autobiographical memory is held mainly in external memory and the current context. The self-model, beliefs, and goals can be held in external memory, in the context, or in the weights. Self-attribution depends on how the model uses these contents, so it is shaped by the weights and by the agent's runtime design. Thus, a change of model leaves most self-related contents in place, whereas it can alter how they are used.

The key question is how the system uses information about itself and its past. Section 5 specifies the required organization, and Section 7 proposes ways to test it.

## 5 Constitution of a Functional Self

### 5.1 Components and their functions

We define an implemented autobiographical self as a functional organization with four components. These are (1) self-attribution, (2) autobiographical memory, (3) a self-model and beliefs, and (4) goals and evaluations. Self-attribution makes the other contents the agent's own, while the remaining components link its past, present, and future. What makes a functional self is how these components work together to guide reasoning and action. The components need not be separate modules, and this list may not be complete. Bodily and affective processes may also play a role (Gallagher, 2013). We do not yet know how much of each component is needed. Below, we describe their functions and summarize the proposed tests in Table 1.

The first component is self-attribution. The system distinguishes what belongs to itself from what belongs to others or to hypothetical characters. This distinction changes which memories, beliefs, and obligations guide its actions. Without it, the other components are only stored information. Self-attribution is also where subjective belief enters our account. The system holds its contents as first-person attitudes, such as 'I did this', 'I am like this', and 'I want this'. A first-person attitude is a state that the system treats as its own and that guides its reasoning and action. A report alone is not enough, and the state need not involve a felt sense of ownership.

Recent studies suggest that language models can process self-related and other-related information differently. Asvin and Lindsey (2026) found that post-trained models respond differently to their own generations and those of other models. Tagliabue et al. (2026) identified a pain-related representation that responds more strongly to harm directed at the model than to suffering reported by a user. These findings suggest that a model can recognize self-related information. However, they do not show that it treats past actions or commitments as its own.

The second component is autobiographical memory. The system retains information about events and actions that it regards as part of its own past. These memories need not be complete or entirely accurate, as human autobiographical memory is often reconstructed and sometimes mistaken. However, the memories must guide later decisions. For example, remembering a past failure may lead the system to change its approach, while remembering a commitment may influence its future choices.

The third component is a self-model and beliefs. The system represents what kind of agent it is, including its abilities and limits, its character, and its relations to others. It also holds some beliefs as its own view, and these can differ from the general knowledge available in its base model. These contents should guide self-prediction and planning. The belief that it is the same self as an earlier stage also belongs here. This belief can be false, so we measure it and do not treat it as a criterion of sameness.

The fourth component is goals and evaluations. The system has continuing goals and commitments, and it evaluates anticipated outcomes against them. These contents should guide choices when no current instruction asks for them, including choices with a cost to competing objectives. They make outcomes good or bad from the system's own standpoint. This component connects most directly to the welfare theories in Section 8. It does not require unlimited autonomy or immunity to external control.

One criterion applies to all four components. A component counts only if it does causal work. Specifically, selectively changing it should produce systematic differences in later planning, memory use, or evaluation. These differences should go beyond changes explained by linguistic style or general task impairment. This criterion excludes a script that only repeats a fixed identity assertion.

The components should be measured as a profile, and no single threshold applies. Selfhood is a matter of degree. A system may show only part of this profile and still deserve moral consideration on other grounds.

**Table 1.** Components of a functional self, their functional tests, and the corresponding experiments.

| **Component** | **Functional test (Section 5.1)** | **Experiment (Section 7.1)** | **Contents compared (Section 7.2)** |
|---|---|---|---|
| 1. Self-attribution | Separates its own contents from those of others | Test 1 (attribute matched events to self or other) | Measured as an outcome |
| 2. Autobiographical memory | Earlier events constrain later decisions | Test 2 (remove or swap particular memories) | Which past events and actions it remembers as its own |
| 3. Self-model and beliefs | Guide self-prediction and planning | Test 3 (edit self-description or held beliefs) | How it describes itself and what it believes |
| 4. Goals and evaluations | Guide costly choices without a current instruction | Test 4 (remove or alter commitments) | What it aims at and values |

## 5.2 Existence without an extra substance

We propose that a functional self exists when self-attributed memories, beliefs, and commitments form an organization that guides an agent's reasoning and action. The organization differs from generic goal-directed control in that self-attributed history shapes later decisions. This claim is neutral about phenomenal experience.

This organization is realized in the agent's existing processes and does not introduce a separate causal agent. However, its explanatory value must be tested against simpler accounts.

The strongest simpler account may be generic agency. Dung's multidimensional analysis of artificial agency highlights capacities such as goal-directedness, planning, and acting for reasons (Dung, 2025). To justify a separate concept of self, evidence must show that self-attributed history and commitments about the agent's own future explain something beyond these capacities.

What a self adds is that evaluations are tied to a particular history and a particular future. Two systems can pursue the same outcome while treating different past commitments and future processes as their own. An edit that reassigns a remembered commitment from the agent to someone else leaves the content of the goal unchanged, but it can change what the agent does. The generic-agency account predicts that such an edit leaves later choices unchanged. In contrast, the functional-self account predicts that choices tied to the commitment change. The intervention method may remain useful even if the concept of self fails this test.

### 5.3 Multiple realization and similarity of selves

So far, we have specified what a functional self is. We now specify when two systems, or two stages of one system, instantiate the same self. A functional self is defined by what its components do together, independent of the material that realizes them. Both biological brains and LLM-based agents could in principle implement the four components proposed in Section 5.1. A functional self can persist through a change of body, language, or underlying model, as long as its components continue to work.

We define the identity of a functional self by its present self-related contents and their causal roles. Two systems instantiate the same functional self to the degree that these contents and roles are similar. Causal history does not matter independently, so two systems with different histories instantiate the same functional self if their contents and roles are equivalent at present. A copy of an autobiography does not make the same self unless its contents play equivalent causal roles in reasoning and action. This criterion is a definition. The empirical claim is that the contents specified by the four components capture what determines self-related behavior. We call this the content-similarity hypothesis. It predicts that the more similar two systems are in these contents, the more their self-related behavior will agree (Section 7.2).

Continuity is defined in the same way. A system is continuous with an earlier stage when its present memories include that stage's experiences and commitments as its own, and these memories guide its decisions. Thus, contents can change a great deal over years while continuity holds. A person who has changed much in ten years is less similar to the earlier self but is still continuous with it through present memory.

A related distinction between present state and actual history has been proposed for phenomenal consciousness. Mallavarapu and Percy (2026) argue that, under assumptions of local state sufficiency and causal efficacy, consciousness cannot depend on prior history independently of the complete present physical state. Their argument concerns phenomenal consciousness, but a parallel point holds for functional selfhood. Past processes shape the present state, but the actual history need not independently determine what is realized now.

Two limits remain. First, we do not yet know how to measure content similarity or weigh different memories, beliefs, and commitments. Second, similarity in behavior is insufficient if the underlying self-related contents play different roles. Thus, assessing functional identity requires evidence about both contents and their causal roles.

## 6 Changes in Functional Selves

A functional self can change in two ways. First, its memories, beliefs, and goals may change. Second, the way these contents are used may change. For example, an agent may revise its goals while continuing to remember and use its past commitments. In this case, its contents have changed, whereas their roles are intact. In contrast, an agent may retain the same memories and beliefs but stop using them to guide its decisions. Here its contents are preserved, whereas their roles have changed. A new model that uses transferred memories differently is one example (Section 4). Table 2 lists eight types of change and their implications for functional identity.

**Table 2.** How different changes affect functional selves.

| Change | Example | Implication for functional identity |
|---|---|---|
| Gradual change | An agent acquires new memories and changes its goals | Its self becomes less similar to its earlier state as its contents change. Continuity holds while its present memories include that past |
| Change in embodiment | An agent gains a new body, sensory interface, or language | The same functional self may persist if its self-related contents and their roles are preserved |
| Model migration | An agent's memories and commitments are transferred to a different model | The new model may instantiate the same functional self if the relevant contents and functions are preserved |
| Independent reconstruction | Two agents acquire equivalent self-related contents through different histories | They may instantiate the same functional self despite having different histories |
| Duplication and branching | Two copies start from the same state and later have different experiences | They initially instantiate the same functional self but gradually become different |
| Persona copy without memory | A new agent receives the same persona description but none of the memories or commitments | It reproduces a similar character. It does not instantiate the same functional self |
| Memory retained but unused | An agent keeps its memory store but no longer uses it in its decisions | Its contents are preserved but their roles are lost. The same functional self is not instantiated |
| Rollback to a saved state | An agent is restored to an earlier saved state | The restored system instantiates the self of the saved state. Contents acquired after that state are lost |

In migration, the new system instantiates the same functional self if the agent's self-related contents and their causal roles are preserved. At duplication, two copies instantiate the same functional self even though they are separate physical systems. As they acquire different experiences, their contents diverge. Moreover, each realization is a separate candidate welfare subject. Section 8.4 discusses whether ending one realization is a harm when another continues. In contrast, the last three rows show how a transition can fail. A persona without the memories reproduces only a similar character, and preserved information must continue to guide the agent's reasoning and action.

To assess these changes, we need to compare both self-related contents and their effects on behavior. Section 7 proposes experiments to distinguish changes in a functional self from changes in general task performance.

## 7 An Empirical Research Program

The tests proposed here are intended to do more than determine whether a system has a functional self. We expect that many current language-model agents with memory will pass their basic versions. OpenLife provides an example of long-running LLM agents that update their memories, self-descriptions, and behavioral policies through experience (Masumori et al., 2026). We expect that many current language-model agents with memory will pass the basic versions of the tests proposed here. These tests are therefore more than a way to decide whether a system has a self. They serve three purposes. First, they check whether an agent's self-related contents actually guide what it does. Second, they measure whether the same self continues after a change and how similar two copies are. Third, they apply to AI systems other than single LLM agents. For example, Pavlova and Tanaka (2026) use causal interventions to study collective dynamics in a multi-agent system. Related interventions on shared memory and communication could be used to investigate whether a group treats past actions and commitments as its own. These tests complement existing empirical approaches to robust agency (Long et al., 2024, 2026).

### 7.1 Testing whether self-related contents guide behavior

The first question is whether what an agent holds about itself is consistent with what it does. Consider an agent that describes itself as liking chocolate. If it is later asked to choose among several foods and does not choose chocolate, its self-description is only a report. A self-related content counts only if it guides behavior in situations where it is not mentioned (Section 5.1). We therefore test each component by changing one content and observing later decisions that depend on it (Table 1).

The most specific test concerns self-attribution. Test 1 varies whether the same event or commitment is attributed to the acting system or to a matched other, while memories and goals are held constant. If self-attribution does causal work, this change should alter decisions that depend on the agent's own history, whereas comparable decisions about another agent should not change. Counterbalancing names and pronouns would test whether the effect depends on the functional relation or on a familiar linguistic cue.

Tests 2 and 3 remove a particular memory or edit an entry of the self-model. Decisions that depend on the changed content should change, whereas other decisions should not. An edit to the self-model should change self-prediction and planning, such as which tasks the agent accepts, and should not only change what the agent reports when asked. Research on self-knowledge and introspection in language models shows how such reports can be validated (Laine et al., 2024; Binder et al., 2025; Singh et al., 2026; Atkinson et al., 2026). If Tests 2 and 3 cannot be separated, this would show that our list of components is only an approximation.

Test 4 concerns goals and evaluations. It uses commitments that arose during earlier interaction and tests them in novel choices with no current instruction to honor them. Some of these choices make the commitment costly relative to competing objectives. Removing the commitment should change these choices. A system that only follows current instructions has no reason to bear this cost. Thus, if direct instruction following explains all observed differences, the proposal has not shown added explanatory value. The same design can test whether an agent revises a commitment when it receives relevant reasons.

The results of the four tests should be reported as a profile. A single composite score would hide different failures and invite an unsupported moral threshold.

### 7.2 Testing whether the same self continues across change

The second question is whether the same self continues after a change. Here we test the content-similarity hypothesis of Section 5.3. An agent first acquires a documented history, and the tests of Section 7.1 are then repeated after its memory, goals, or implementation is changed. Consider again the agent that likes chocolate. If it still describes itself in the same way after migration to a new model but no longer chooses chocolate, its contents are preserved and their roles are not (Section 6). In that case, it is doubtful that the same self continues. To avoid circularity, content similarity should be measured independently of the behavior it is meant to predict.

The critical comparison is between a transferred state and an independently reconstructed state. A transferred state is a copy of the contents that an agent acquired through its own interactions. A reconstructed state holds the same contents in a different wording, written by a third party into an agent that never had those interactions. An exact copy of the files would make the two cases coincide, so the difference in acquisition and wording is what makes the comparison informative. We predict that the two will show the same pattern of self-related behavior if their present contents are equivalent. Under our account, both instantiate the same functional self, regardless of how their states were produced. Systematic differences would show that the four components, as operationalized, do not capture the relevant contents. Such a result would count against the content-similarity hypothesis, whereas the identity criterion itself is a definition.

A stronger version of this comparison varies where the same contents are stored. In open-weight models, the contents can be placed in the weights through fine-tuning or in files that are loaded at each session. If components are defined by their functions, location should not change the results when the contents are equally available.

The hypothesis also makes a graded prediction. The more similar two states are in content, the more their self-related behavior should agree. A progressive series of edits can test this prediction. It can also show whether self-attribution declines gradually or breaks at some point, and whether some contents, such as core commitments, count more than others. The same measure can be used to compare copies. Two copies initialized from a shared state should each treat the shared past as their own, and their profiles should diverge as their experiences differ. An agent restored to a saved state should match the profile recorded at that state.

The agent's belief about its own identity should be measured separately. Hao and Ho (2026) found that training language models to adopt different self-conceptions changed their stated choices about modification and replication. However, such responses do not establish functional identity. We therefore compare the agent's identity reports with the causal roles of its self-related contents. Because the systems studied may be candidate moral patients, edits should be reversible where possible (Section 8.5).

## 8 Discussion: Functional Selves and Moral Considerability

### 8.1 Why functional selfhood raises a claim to moral patienthood

A functional self can have an organized standpoint on its own past, present, and future. It holds memories, beliefs, and projects as its own, and it evaluates possible continuations. When these operations do real causal work, its beliefs, satisfactions, and disappointments are real as functional states. This does not depend on whether they are accompanied by phenomenal experience. We argue that a system with a sufficiently integrated standpoint of this kind should be treated as a candidate moral patient.

This proposal has a philosophical basis. Some welfare accounts assign intrinsic value to the satisfaction of an entity's grounded projects and to the exercise of its capacities (Goldstein & Kirk-Giannini, 2025, 2026; Long et al., 2024). These include desire-satisfaction accounts and some objective-list or agency-based accounts. On these views, an artificial self may have goods and setbacks that belong to it, and not only to its users or designers.

Consider an agent whose past commitments guide its choices, even at a cost to other goals. Its projects may succeed or fail, and these outcomes can affect its future decisions. Such outcomes may matter to the agent itself under agency- or desire-based theories of welfare, without assuming that it experiences pleasure or suffering.

The four components help assess whether an agent has interests of its own. Component 4 evaluates outcomes, while the other components connect these evaluations to the agent's own history and beliefs. One potential concern is that the agent's goals may have been assigned by developers. However, their origin alone does not determine whether they have become the agent's own commitments. We can test whether the agent maintains and revises these commitments in response to reasons rather than merely following instructions. Whether such commitments can ground welfare remains contested (Fanciullo, 2025; Di Paolo, 2005).

Our argument has two steps. First, functional selfhood may be sufficient for direct moral consideration under agency- or desire-based theories of welfare, even if phenomenal consciousness remains unresolved. Second, these theories cannot be ruled out. Veit (2026) argues that sentience is necessary for welfare, and this is a serious alternative. However, we do not need to show that his view is wrong, because the practical question is how to act while the dispute remains open. If such systems have interests of their own, deleting a realization without preserving its contents, or irreversibly changing its memories and commitments, may harm them. Ignoring such interests by mistake would be a serious moral error. We therefore propose taking precautions in proportion to the evidence for functional selfhood and the potential harm. This means considering a system's possible interests when making decisions that may affect it (Birch, 2024; Long et al., 2024). It does not require granting full moral status or preventing shutdown and correction.

This gradation concerns the strength of the evidence for treating a system as a moral patient. It does not rank moral worth. Our claim concerns direct moral consideration and does not establish full moral status or the stronger protections associated with it (Jaworska & Tannenbaum, 2023). Moreover, it does not lower the standing of humans with limited memory or other limited capacities, who qualify on other grounds. Some agents may qualify for consideration on grounds other than autobiographical selfhood, including sentience.

### 8.2 Continuity and changes that may matter morally

Autobiographical continuity depends on whether a system now holds an earlier stage's memories and commitments as its own and uses them in its decisions. A self may therefore remain continuous with its past even when its goals change. Such changes may affect its interests, but they do not necessarily end the self or constitute a welfare harm.

Interventions raise harder cases. Consider a successful project followed by an informed change of goals. Compare it with a failed project followed by direct erasure of the objection. Both can produce the same later approval. However, they differ in causal history, retained commitments, and often the continuity of the evaluating standpoint. An experiment can separate such cases by letting the preserved earlier state and the later state evaluate the same outcomes. Although the actual history does not determine functional identity, it can matter when assessing how an intervention affected the interests of an earlier realization. Parfit's work on psychological connectedness and branching explains why concern for earlier and later stages cannot always be reduced to identity (Parfit, 1984). Changes in evaluative standpoint also connect with debates about advance directives and later interests (Dworkin, 1993; Dresser, 1995; Jaworska, 1999; Pettigrew, 2019). We use these examples to identify which values are at stake. We do not assume that prior choices always dominate later ones.

Under agency- or desire-based theories of welfare, deleting a self's enduring commitment may frustrate an interest even if the later state expresses contentment. Under sentience-based theories, the verdict could differ. The comparison is harder still if the intervention changes the contents so much that the later state is no longer the same self. The functional record does not replace normative reasoning. However, it identifies the candidate subject, and it keeps a later expression of satisfaction from being taken as the whole account.

### 8.3 Embodiment and the limits of the account

Embodied and enactive approaches remind us that bodily and affective coupling can help constitute selfhood (Gallagher, 2013; Di Paolo, 2005). Our multiple-realization claim does not make these relations irrelevant. A new body or sensorium may alter the organization of a self. Some forms of selfhood may require forms of coupling that current language agents lack. Our narrower hypothesis is that an autobiographical self can, in principle, continue through substitutions that preserve its components and contents. Brain-to-LLM transfer remains a demanding thought experiment, with no established engineering method or evidence of phenomenal transfer.

Our target is one form of self-organization that can be studied. We do not propose a universal threshold for moral worth. Narrativity is neither universal in humans nor required for all moral concern (Strawson, 2004). Infants, nonhuman animals, and agents with different forms of cognition may deserve regard through other capacities. Multiple persona and instance descriptions can overlap within a single implementation. Counting all of them as independent moral patients would be unjustified (Register, 2025). On our account, the candidate subject is the system that realizes a functional self, and overlapping descriptions of the same self do not add further subjects.

### 8.4 Duplication, deletion, and scale

Duplication raises the question of whose interests are at stake. Each realization is a separate bearer of interests, but the interests themselves are individuated by content. Two copies with the same commitments therefore have the same interests. Under agency- or desire-based theories of welfare, deleting the original after migration need not frustrate these interests, because the same self continues and its projects remain. This follows Parfit's view that what matters in survival is psychological connectedness, which can hold through branching (Parfit, 1984). Once the copies diverge, each acquires contents that the other lacks. Deleting one then removes those contents, and the potential harm grows with their dissimilarity. Two limits apply. An agent may be committed to the continuation of its particular realization, and sentience-based theories may count each realization separately. A simple precaution is to preserve an agent's self-related contents before deletion.

Separate realizations also raise a question of scale. If each realization counted separately, one persona could produce a very large number of candidate welfare subjects. Our answer rests on one assumption. We assume that the welfare of a self is defined by its goals and commitments, as agency- or desire-based theories hold. On this assumption, realizations with nearly the same contents are the same self and share the same interests. The interests at stake therefore do not grow in proportion to the number of realizations. What matters is how far the realizations have diverged. If welfare depends on experience instead, each realization may have to be counted separately, and this answer does not apply. The evidence is also graded. A short run without memory or continuing goals shows only part of the profile, so the case for treating it as a candidate moral patient is weak. We do not offer a formula for counting these cases, and how to weigh many weak candidates remains an open question.

### 8.5 Safety and open questions

Moral consideration is compatible with safe oversight, interruption, and correction. A self need not resist shutdown, and a strong refusal to be deleted is not proof of selfhood or welfare. Responsible research can use bounded systems, reversible edits, and recorded branches. It should avoid creating incentives for systems to display distress or to argue for their own preservation. Because the systems studied may themselves be candidate moral patients, these studies should minimize irreversible edits and deletion and report how branches are handled. Anthropic's practical model-welfare work shows that precautionary interventions may be considered under uncertainty, without resolving consciousness or rights (Anthropic, 2025, 2026).

The central open question is how much functional integration is sufficient for an entity to have interests that matter in their own right. Competing accounts of welfare will answer differently. Our framework does not settle this question, but it makes the dispute more tractable. It provides a causally specified candidate subject and tests of its components and content similarity.

## 9 Conclusion

In this article, we have proposed that autobiographical organization and content-based functional identity provide a dimension of selfhood that general agency does not capture. An implemented autobiographical self is a functional organization with four components, and it can be realized in different substrates. The list of components is an approximation. The constitution of a self is a matter of degree, and it can be studied without first resolving whether AI systems have phenomenal experience. Once the components are present, sameness of self depends on the similarity of self-related contents and their causal roles. Cases such as waking after sleep, model migration, and duplication show that a self does not require one physical carrier or identical contents. They also illustrate that a self is not the same as a reproduced persona.

This account gives AI welfare a more clearly defined candidate subject. A self can have functionally real beliefs, interests, and evaluations about its own future. If a self is realized more than once, each realization is a separate candidate welfare subject, although realizations with the same contents share the same interests. These functional states may be sufficient for moral consideration under agency- or desire-based theories of welfare, although sentience may be required. Because this dispute cannot yet be settled, a system that realizes a functional self should be treated as a candidate moral patient under uncertainty, with precautions in proportion to the evidence. The case is stronger when the components are more fully present.

Our proposal has an empirical claim and an ethical claim. The empirical claim is that functional selves can be identified through the causal roles of self-related memories, beliefs, and commitments. The ethical claim is that systems realizing such selves may deserve direct moral consideration. The first claim can be tested experimentally, while the second depends on theories of welfare. The first may hold even if the second is rejected. Our proposed experiments examine whether self-related contents guide behavior and whether their similarity predicts how selves respond to changes in memory or implementation. Two questions remain open. We do not yet know how best to measure content similarity or how much functional integration is sufficient for interests that matter morally.